# Large orbital magnetic moment in VI$_3$


*Dávid Hovančík[1]\*, Cinthia Piamonteze[2]\*, Jiří Pospíšil[1], Karel Carva[1], Vladimír Sechovský[1]*

[1] Charles University, Faculty of Mathematics and Physics, Department of Condensed Matter

Physics, Ke Karlovu 5, 121 16 Prague 2, Czech Republic

[2] Swiss Light Source, Paul Scherrer Institut, CH-5232 Villigen PSI, Switzerland



Abstract: The existence of the V$^{3+}$- ion orbital moment is the open issue of the nature of

magnetism in the van der Waals ferromagnet VI$_3$. The huge magnetocrystalline anisotropy in

conjunction with the significantly reduced ordered magnetic moment compared to the

spin-only value provides strong but indirect evidence of a significant V orbital moment. We

used the unique capability of X-ray magnetic circular dichroism to determine the orbital

component of the total magnetic moment and provide for the first time a direct proof of an

exceptionally sizable orbital moment of the V$^{3+}$ ion in VI$_3$. Our ligand field multiplet

simulations of the XMCD spectra in synergy with the results of DFT calculations agree with




the existence of two V sites with different orbital occupations and therefore different OM magnitudes in the ground state.

**Introduction**

Two-dimensional (2D) van der Waals (vdW) magnetic materials have recently attracted ever-increasing interest from materials researchers because their stable magnetic ordering, even in atomically thin monolayers, provides great opportunities for spintronic devices[1-5]. The magnetic ordering in 2D relies on magnetic anisotropy since Mermin and Wagner[6] have shown that long-range magnetic order cannot exist in the one- or two-dimensional isotropic Heisenberg model. As proposed by Bruno[7] the magnetocrystalline anisotropy energy is proportional to the orbital moment (OM). In the extensively studied $CrI_3$, the electronic configuration of the $Cr^{3+}$ ion which experiences an octahedral crystal field is $t_{2g}^3$, $S = 3/2$, $L$ = 0. The small OM value below 0.1 $\mu_B$ was confirmed by XMCD measurements[8]. As proposed by Kim *et al.*[9], the magnetic anisotropy of $CrI_3$ comes from the iodine $p$ orbital spin-orbit coupling (SOC) and the *p-d* covalence between Cr and I, hence, being relatively small. Here we focus on a Mott insulator $VI_3$, which began to be intensively investigated after



the papers of Son *et al.*[10], Kong *et al.*[11], and Tian *et al.*[12] on ferromagnetism in bulk crystals were published. The recent work of Lin *et al.*[13] showing an anomalous increase of $T_C$ for one or few monolayers of $VI_3$ further boosted this interest. In $VI_3$ the electrons in a partially filled $t_{2g}$ level may possess an effective OM l = 1[14], which for the occupancy of two electrons leads to S=1, L=1 configuration. As S and L are antiparallel due to the less than half-filled V *d*-shell, one would expect a total magnetic moment of 1 $\mu_B$, if the picture is not affected by other effects quenching the orbital moment. Early publications on $VI_3$[10-12] showed huge disagreement on the magnitude of the out-of-plane ($c^R$-axis; $c^R$ stands for *c*-axis in rhombohedral structure) magnetic moment (at 2 K and 5 T the values 2.5, 1.25, and 1.1 $\mu_B$/V, respectively). Later, Liu *et al.* reported a moment of 1 $\mu_B$/V and proposed an unquenched OM[15]. This scenario was further supported by neutron diffraction experiment done by Hao *et al.*[16], which found a magnetic moment of 1.2 $\mu_B$/V at 6 K. Electronic structure calculations have found a ground state where the $e'_g$ orbital is half occupied leading to an unquenched OM[17,18]. Strong correlations represented by Hubbard U are sufficient to open a gap between SO split bands, so that this solution is semiconducting. On the other hand, another calculation suggests different semiconducting ground state with fully occupied $e'_g$ orbitals[19], which leads to a negligible orbital moment. Direct evidence and quantification of an unquenched OM is still lacking.



$TX_3$ transition metal trihalides ($T$ = transition metal, $X$ = Cl, Br, I) adopt two common layered crystal structures with $T^{3+}$ ions arranged in a honeycomb network at the edge-sharing octahedral coordination by six $X$ ions (see Figure 2a). $VI_3$ undergoes a structural transition at 79 K between the high-temperature rhombohedral $R\overline{3}$ and low-temperature monoclinic $C2/m$ variant[11,20-23]. Below 50 K it orders ferromagnetically with an easy-magnetization direction tilted by ~ 40° from $c^R$-axis[16,24]. A strong magnetic anisotropy is observed, which reinforces the proposition of an unquenched OM[25]. Around 36 K a second magnetic phase transition was proposed separating the states with one (at temperatures between 36 and 50 K) and two inequivalent magnetically ordered V sites (at lower temperatures)[26]. In addition, a symmetry lowering from monoclinic to a triclinic structure appears at 32 K upon cooling[23,27]. Magneto-transport measurements also show a qualitatively different ferromagnetic state below 40 K, compared to the one between $T_c$ and 40 K[28].

Motivated by the multiple hints for non-zero OM in $VI_3$ we utilized the unique ability of X-ray magnetic circular dichroism (XMCD) to determine spin and orbital moments separately [29,30] in an element-specific fashion. Our results confirm the existence of an exceptionally large OM. Ligand field multiplet simulations of the XAS and XMCD spectra are in qualitative agreement with the existence of two V sites with different orbital occupations and therefore different OM at 2 K.



**Results and discussion**

Magnetization measurements show a 1.23 $\mu_B$/V-ion along the easy magnetization direction, well matching the neutron-experiment result[16] (see SI, Figure S1). Angular dependence of 5-T magnetization in the $ac^R$-plane at 2 K shown that the easy axis which is tilted $\sim 40°$ from the $c^R$-axis, also in agreement with previous publications[16,24,31].

Figure 1a shows the V $L_{3,2}$ ($2p_{1/2,3/2} \rightarrow 3d$, between 505 and 530 eV) helicity-averaged XAS of $VI_3$ taken at 300K and 2K. The spectrum agrees well with the $V^{3+}$ valence state published for $V_2O_3$ and the previous XAS published for $VI_3$[32,33]. For the 2K spectrum besides the V $L_{3,2}$ absorption edges, we detected another two resonances above 530eV which were identified as the O K absorption edge ($1s \rightarrow \pi^*$, $1s \rightarrow \sigma$) of the $O_2$ molecule[34] see Figure 1b. The presence of O K edge XAS of different intensities was observed at each temperature below 90 K. We ascribe this effect to the finite amount of frozen $O_2$ ($T_{boiling} < 90K$) coming from the chamber residual pressure. The 90 K XAS measurement (see SI, Figure S2) shows an oxygen-free spectrum measured after the low-temperature data, clearly showing that the $O_2$ is not intrinsic to the sample surface.



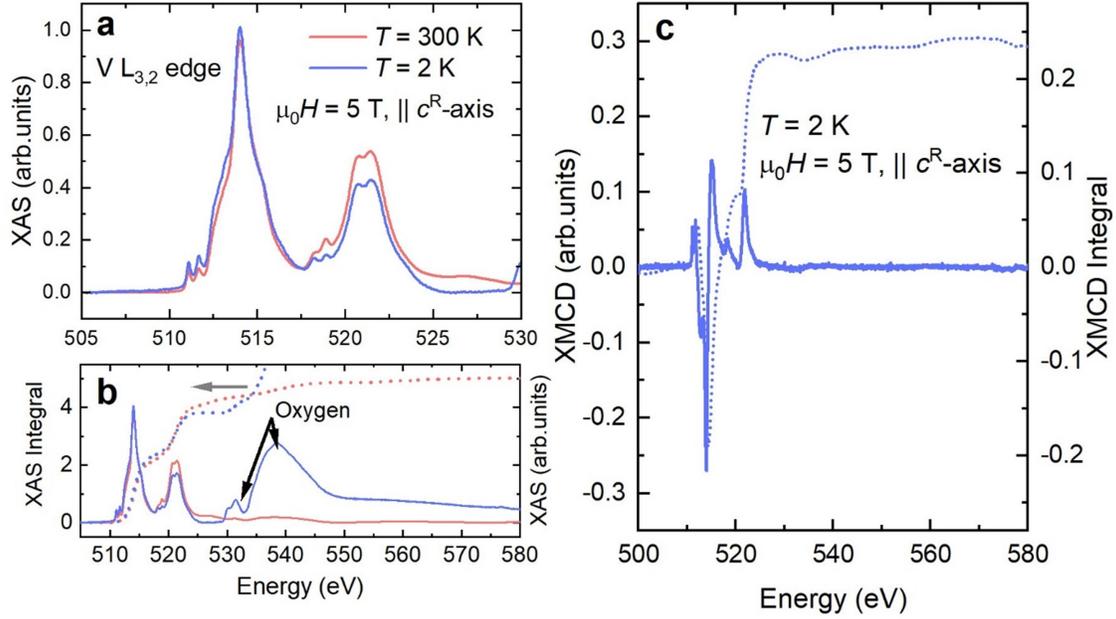

Figure 1. a) V $L_{2,3}$ XAS spectra of $VI_3$ measured at 300 K and 2 K after the removal of the step function (continuous lines). b) The XAS spectra in extended energy region and corresponding XAS integrals (dotted lines). c) XMCD spectrum at 2 K (continuous line) and corresponding integral (dotted line).

Figure 1c shows the V $L_{3,2}$ XMCD spectra derived as $\sigma^+(\omega) - \sigma^-(\omega)$ measured at 2K and 5T applied parallel to the $c^R$-axis. From the XMCD integral shown in figure 1c, one can notice that the value of the integrated intensity is finite and positive. According to the magneto-optical sum rules of the XMCD spectra[29,30] (see SI, Eq (S1)) the total integral of the XMCD signal is proportional to the orbital magnetic moment. Therefore, the positive finite integral undoubtedly confirms the nonzero out-of-plane ($c^R$-axis) OM component antiparallel



to the spin moment, as expected. Applying the magneto-optical orbital sum rules to the spectra we obtain $m_{orbital}$ = 0.6 (1) $\mu_B$. The error bar takes into account the uncertainty in determining the XAS integral due to the oxygen contribution, as detailed in the SI. The spin sum rule cannot be applied to V spectra due to the large overlap of states from the $L_3$ and $L_2$ edge, which makes the calculation of a correction factor unfeasible[35,36].

In VI$_3$ a trigonal distortion leads to a splitting of $t_{2g}$ level into $a_{1g}$ singlet and $e'_g$ doublet. Two possible ground states for $3d^2$ configurations are: (i) lower energy of $e'_g$ doublet which would lead to the ground state with a fully occupied $e'^2_g$ level and orbital singlet with L= 0; (ii) electronic occupation of $a^1_{1g}\,e'^1_-$ levels where the $e'_g$ doublet is further split by spin-orbit interaction (see Figure 2 b, c and d). In the second case, the electron in the $e'_-$ level can be assigned an effective moment analogous to $p$ electrons, l = 1, and the ground state may be in the high OM state with L= 1.



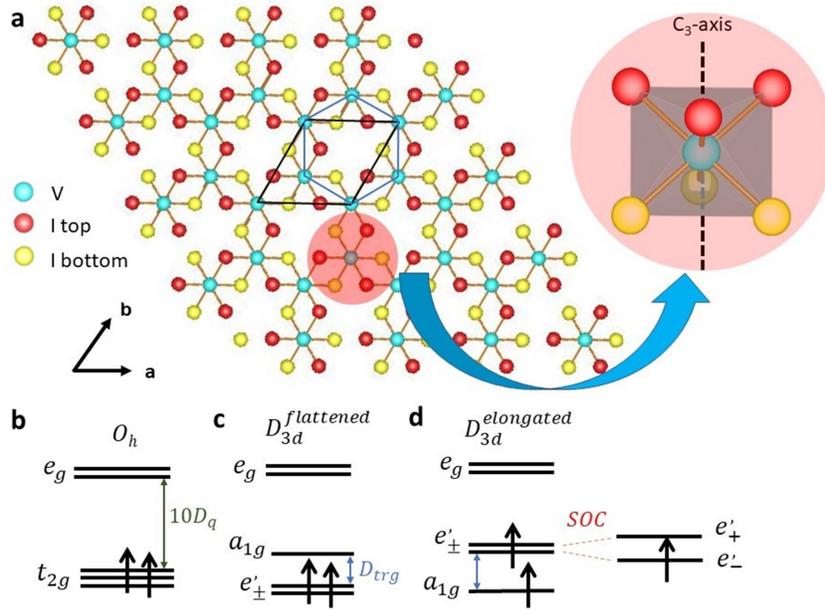

Figure 2. a) Crystal structure of VI$_3$ monolayer. The VI$_6$ cluster is emphasized. b) c) and d) Crystal field splitting for O$_h$ and D$_{3d}$ (flattened/elongated octahedra) symmetry and corresponding electron occupation.

In the case of a trigonally distorted cubic lattice, one can decide which of the two above cases is energetically favorable knowing whether the cubic body diagonal (along C$_3$- axis) is elongated or shortened. In VI$_3$ the symmetry of more distant neighbors of V beyond the octahedral I cage differs significantly from the octahedral one (see Fig. 2), which may lead to corrections (of trigonal symmetry) comparable to that of the small I cage distortion. Therefore, we have employed the electronic structure DFT (for details see SI, DFT) calculations to get a more accurate picture. When both, the spin-orbit interaction and correlation effects in terms of Hubbard U are included, these calculations converge to two



strikingly different solutions: either a state with quenched OM, typical for 3$d$ transition metals, or a state with an exceptionally high OM[17,18]. The latter state is energetically favorable, but the energy difference is small, approximately 5.6 meV according to our calculation. Under some circumstances, the state with quenched OM may become preferred. For the spin moment, all calculations predict values close to 2 $\mu_B$ in agreement with Hund rules.

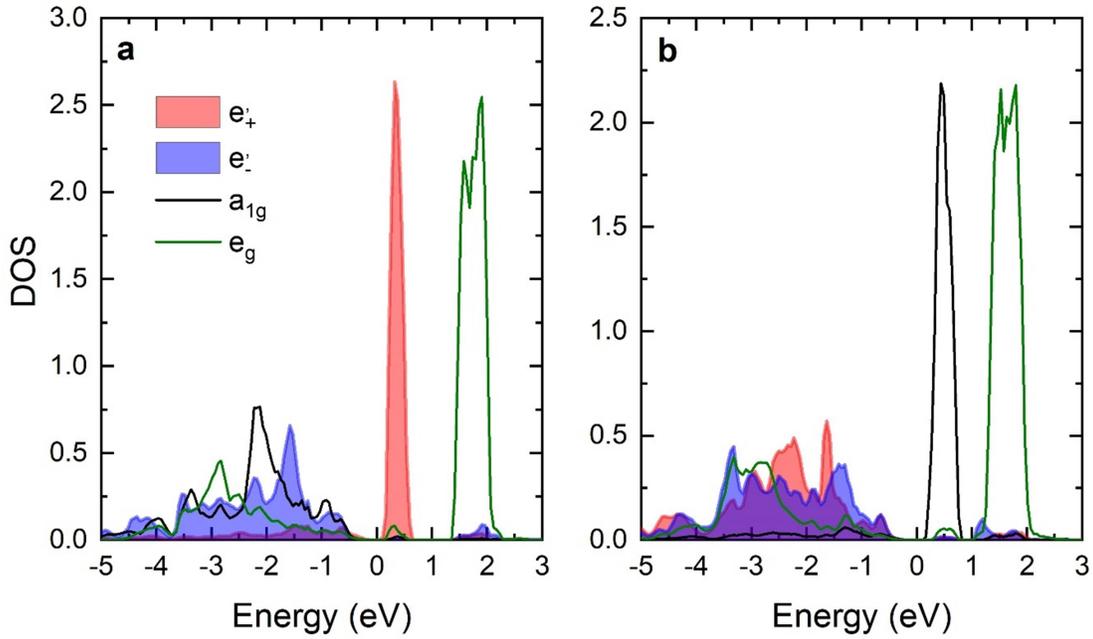

Figure 3. Calculated orbital-resolved DOS of VI$_3$ for majority spin V 3d electrons. a) solution with high 3$d$ OM value b) solutions with quenched OM.

For the solution with high OM, the orbital-resolved DOS for the majority spin (Figure 3 a) shows that the $e'_-$ states are almost fully occupied while $e'_+$ states are empty. Notably the relevant bands are rather broad and $a_{1g}$ character bands have a different evolution in



k-space than $e_g'$ bands. Nevertheless, one can say that $a_{1g}$ states are generally energetically lower than $e_g'$ as expected for this situation. On the other hand, the solution with the quenched OM shows a complete occupation of $e_-'$ and $e_+'$ states whereas, the $a_{1g}$ orbital states are above the Fermi level. This solution was found previously using the VASP code with a smaller Hubbard U = 2 eV[19].

To help understanding of the V ground state we have also performed ligand multiplet simulation of the XAS and XMCD spectra incorporating the strong interaction between the $2p$ core hole and $3d$ electrons[37]. Given the broadness of some of the bands, the use of an *ab initio* electronic structure to obtain parameters for the multiplet calculation has to be done with caution. From the calculation performed without Hubbard U in order to look at single electron energy levels we estimate $E(t_{2g}) - E(e_g)$ splitting $10D_q$ to be in the range of 1.3 - 1.8 eV. For the multiplet simulation, we have used $10D_q$ = 1.5 eV from this range. Other parameters and details of the simulation are described in the SI.



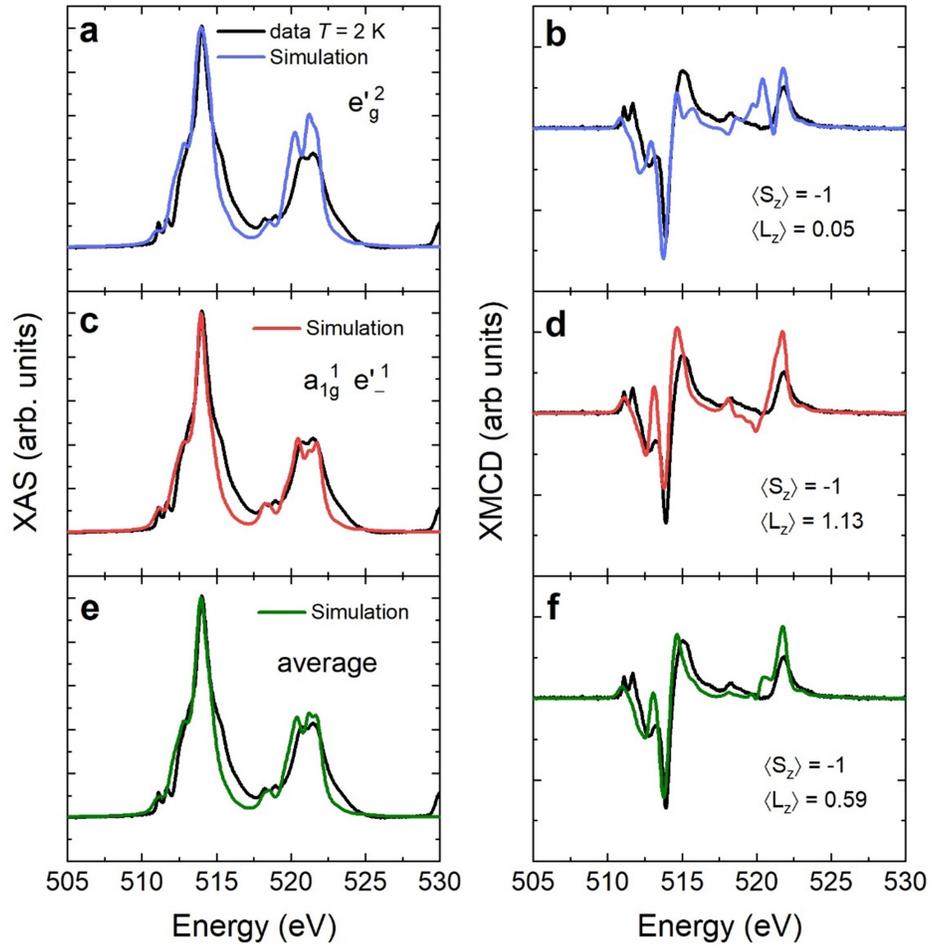

Figure 4. X-ray absorption (left panel) and x-ray magnetic circular dichroism (right panel). The data is plotted in black while the simulations are in blue, red, and green. a) and b) simulations corresponding to $D_{trg} = -0.15\ eV$ and a quenched OM. c) and d) simulations corresponding to $D_{trg} = 0.3\ eV$ and an unquenched OM. e) and f) XAS and XMCD simulation is the average of (a) and (b) simulations which corresponds to an OM of ~ 0.6 $\mu_B$.



Figure 4 shows the measured spectra compared to simulations. Blue curves (Figure 4 a, b) correspond to a negative $D_{trg}$ energy splitting which leads to $e_g'$ orbital as the lowest in energy. This simulation corresponds to $\langle S_z \rangle = -1$, $\langle L_z \rangle = 0.05$ for V in the ground state at 10 K. The red simulation (Figure 4 b, c), on the other hand, corresponds to a positive $D_{trg}$ of the $t_{2g}$ levels with $a_{1g}$ as the lowest energy orbital giving $\langle S_z \rangle = -1$, $\langle L_z \rangle = 1.13$. The largest discrepancy for the simulation with $e_g'^{\,2}$ is an additional positive peak at the $L_2$ XMCD around 520 eV. The $a_{1g}^{\,1} e_-'^{\,1}$ ground state simulation shows a different intensity ratio compared to the data for the positive and negative peaks at the $L_3$-edge XMCD between 513 and 515 eV. The $a_{1g}^{\,1} e_-'^{\,1}$ simulation gives an unquenched OM, in qualitative agreement with the experimental finding, however, the size of $m_{orbital}$ is overestimated by a factor of 2, approximately. In the inelastic neutron scattering experiments, Lane *et al.*[38] have found that their observations fit only with a system composed of two simultaneously coexisting V sites with opposite trigonal distortion, where one leads to the quenched OM while the other one leads to an unquenched OM. Various kinds of coexisting different stable states for V have already been suggested in other works[17,26,39]. The $d$ SOC and crystal field parameters used for the simulations shown in Figure 4 are in close agreement with those used by Lane *et al.*. To simulate the co-existence of two V sites with opposite distortions we take the average (50% occupancy each) of the simulations for $e_g'^{\,2}$ and $a_{1g}^{\,1} e_-'^{\,1}$ ground state which is represented by the green curve in



Figure 4 e, f. The qualitative agreement between data and simulation is improved in comparison to the single site simulations. Most importantly, the average OM from the two sites, which corresponds to 0.59 $\mu_B$ is in very good agreement with the value obtained from the XMCD orbital sum rule. Therefore, our results support the picture of two coexisting inequivalent V magnetic sites.

Recently an ARPES investigation has revealed a significant occupation of $a_{1g}$ orbitals, in addition to $e'_g$ orbitals[33]. In their work, the authors attribute that to $V^{2+}$ at the surface, although their XAS agrees with $V^{3+}$. The argument for this proposition is based on a prediction that $e'_g$ should be fully occupied at the ground state and only the existence of $V^{2+}$ could then explain the observed $a_{1g}$ occupation. The existence of a ground state with the $a_{1g}$ orbital occupied and a high OM, as found here following previous works[17,18,38] would be a reasonable explanation of the ARPES results.

The importance of both many-body effects as well as solid-state hybridization, and the possible presence of multiple V sites in this system represent a challenging task for the theory; a more accurate description of the spectra may be achieved with the use of advanced configuration interaction techniques or the Bethe-Salpeter equation[37,40].

## 4. Conclusions



In summary, our XMCD results unequivocally demonstrate the existence of an unquenched OM of V in VI$_3$, thus resolving the long debate on this issue. Our findings are connected to theoretical models of the V ground state in VI$_3$, help us to reveal which orbitals are occupied, and explain the large magnetic anisotropy observed in this system. Using ligand field multiplet simulations, we could show that the ground state of the V ion would agree with inelastic neutron scattering results where two V sites with opposite trigonal distortion were proposed.

## Acknowledgments


This work is a part of the research project GAČR 21-06083S which is financed by the Czech Science Foundation. The experiments were carried out in the Materials Growth and Measurement Laboratory MGML (see: http://mgml.eu) which is supported by the program of Czech Research Infrastructures (project no. LM2018096). This project was also supported by OP VVV project MATFUN under Grant No. CZ.02.1.01/0.0/0.0/15_003/0000487. We thank A. Marmodoro for valuable discussions. This work was supported by the Ministry of Education, Youth and Sports of the Czech Republic through the e-INFRA CZ (ID:90140).

# Supplementary Information -



The single crystals of $VI_3$ were prepared by the CVT method directly from the stoichiometric ratio (1:3) of elements. All elements as well as the filing of the quartz ampoule were performed in a glove box under Ar inert atmosphere. The degradation process of $VI_3$ was described in detail by Kratochvilova et al.[1]. The chemical composition of single crystals was verified by scanning electron microscopy (SEM) in combination with energy-dispersive X-ray detector (EDX).

MAGNETIZATION MEASUREMENTS

We performed detailed magnetization measurements of $VI_3$ bulk samples at 2 K to check the total value of V moment. The magnetization data was measured in magnetic field up to 7 T using a SQUID magnetometer MPMS7T (*Quantum Design Inc.*). To probe the angular dependence of magnetization in $ac^R$-plane (at 2 K in 5 T) we used a homemade rotator with rotation axis orthogonal to the applied magnetic field. The angular-depended data was measured in relative units and then scaled to the data measured along the $c^R$-axis and $ab$-plane.



The results are shown in Figure S1. The maximum magnetization value 1.23 $\mu_B$ was found along direction tilted by ~ 40º from the $c^R$-axis.

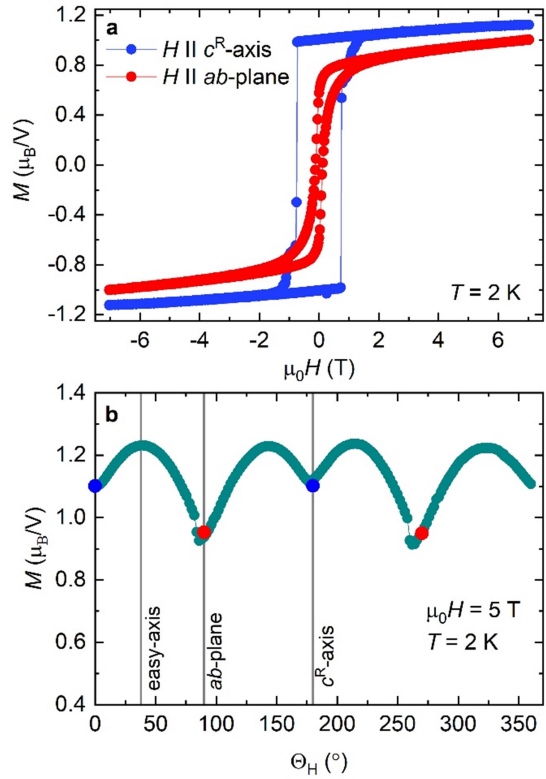

**Figure S1** - top: magnetization isotherms measured at 2 K bottom: angular dependent magnetization. The angular dependence was scaled according to the absolute values of magnetization measured along $c^R$-axis and $ab$-plane (see blue and red dots).

**XMCD sum rules**



We used magneto-optical sum rules[2,3] to analyze the direction and value of the projection of an orbital magnetic moment ($m_{orbital}$). The spin sum rule requires a separation of the XMCD contribution from the $L_3$ and $L_2$ edges. Correction factors of the spin sum rule have been calculated for the late transition metals[4,5]. However, for Ti, V, and Cr the overlap is large and no correction factor can be properly calculated. Therefore, we focus only on the orbital sum rules given by the expression below:

$$\widehat{\boldsymbol{P}} \cdot \boldsymbol{m}_{orbital} = -\frac{4}{3}\frac{q}{r}n_h\mu_B \qquad\qquad (1)$$

where $\widehat{\boldsymbol{P}}$ is the x-ray propagation unit vector, $q$ and $r$ are defined in figure S2, $n_h$ stands for the number of holes per atom in the $3d$ orbital, $\mu_B$ is Bohr magneton. In expression (1) the quantity $r$ is calculated from the XAS defined as the sum of individual spectra measured with left and right circularly polarized X-rays.



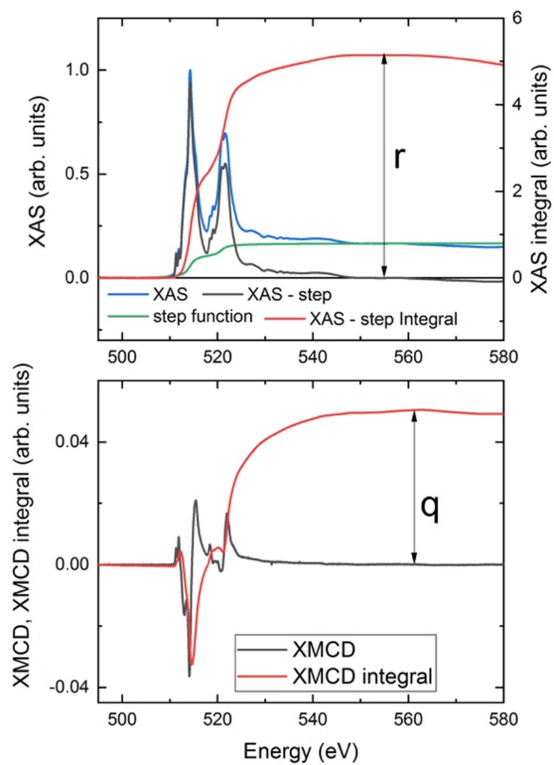

**Figure S2**- top: XAS and bottom: XMCD measured at 90 K normal incidence, 5 T applied field parallel to the X-ray direction.

Figure S3 shows the XMCD, and corresponding integral measured at 2 K, 5 T, normal incidence. The integral value indicated is calculated as the average in the energy region 540-580eV and the error bar the standard deviation. This data was used for the OM calculation presented in the manuscript.



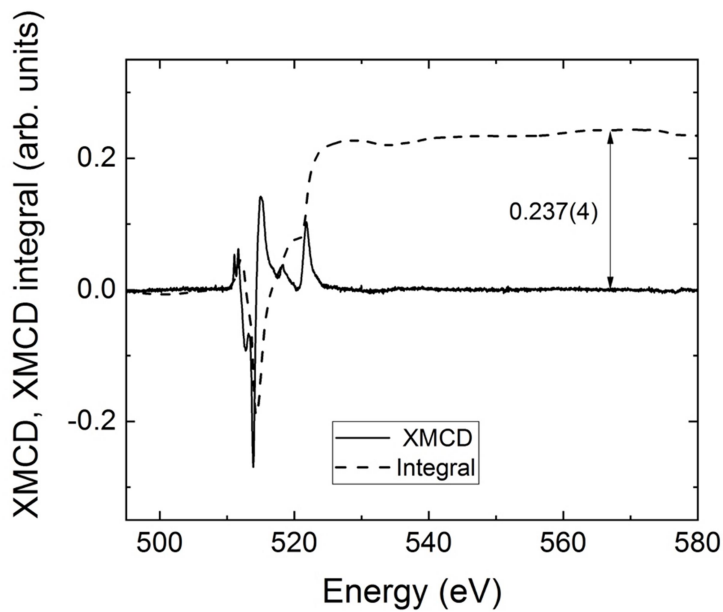

**Figure S3** - XMCD measured at 2 K, 5 T normal incidence and corresponding integral. The integral value indicated is the average within the energy range 540-580 eV and the error bar is the standard deviation.

As discussed in the manuscript, the XAS measured at low temperature presents an additional contribution of the oxygen K-edge, which makes difficult a direct calculation of the XAS integral. There are different ways estimating the XAS integral for the sum rule, which we discuss in sequence. Figure S4 shows the XAS spectra and corresponding integral for the 2 K and 300 K data. One way to calculate the XAS integral is simply to adjust the step function to the 2K XAS spectrum around 527 eV subtract the step function from the XAS and take the integral at this energy. The value of the XAS integral is 3.76 and the corresponding OM, if we



use the XMCD integral from figure S3 and $n_h = 8$, is 0.67 $\mu_B$. The issue with this method is that normalization of the step function cannot be done at a high-energy post-edge as it is usually done.

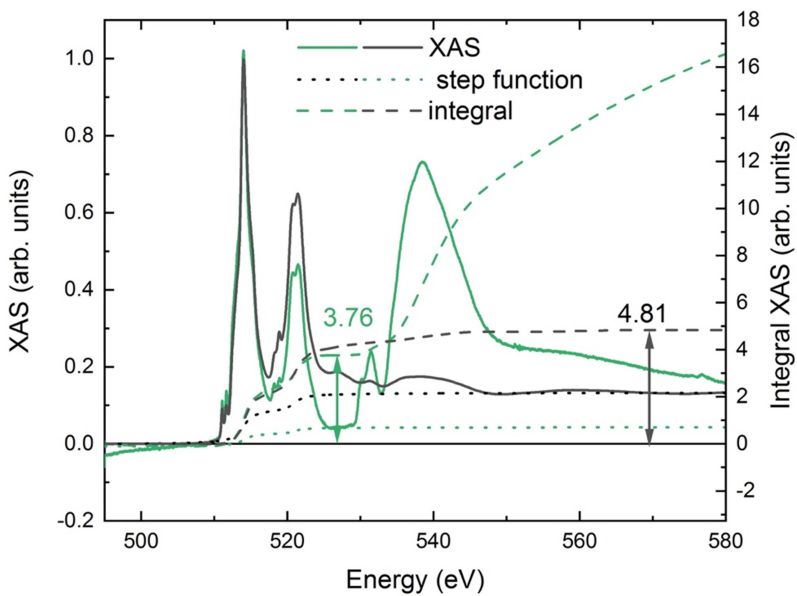

**Figure S4** - XAS (continuous line), step function (dotted line) and corresponding integral after step function removal (dashed line) for the spectrum measured at 2 K (green) and 300 K (dark gray) in normal incidence. The value of the respective integrals is given above the arrows at the energy they were taken.

A second method would be to take the integral of the data measured at 300 K, where the oxygen spectrum does not interfere, and the step function can be normalized at high energy. In this case the step function is normalized around 570eV and the corresponding integral after the



step function removal is 4.81. The resulting OM is 0.53 $\mu_B$. This method is better than the previous one because we can calculate the full XAS integral more precisely. The disadvantage is that the XAS at 300K is not identical to the one at 2K. As seen in figure S4, the relative ratio of $L_3$ and $L_2$ peaks is different.

For this reason, we suggest a third method, which is to estimate the difference in the XAS integral, if the step function is normalized at 527 eV or at 570 eV and use that correction factor to estimate the correct XAS integral for the 2 K spectrum with the step normalized at 570 eV. The correction factor is calculated using the 300 K data. As shown in figure S5 the XAS integral of the 300 K data taken at 527 eV, with the step function normalized to the data at this energy is 3.53. Therefore, the integral taken just above the $L_2$ edge corresponds to 3.53/4.81*100=73% of the integral taken at 570 eV. Using this relative difference between the 300 K integrals at two different energy points we can attempt to correct the 2 K XAS integral which can only be taken in a restricted energy range. We obtain: 3.76/0.73 = 5.12 as an estimate for the 2 K XAS integral value if there would be no Oxygen contribution to the spectrum. If we use this value for the XAS integral, we obtain an OM of 0.49 $\mu_B$.

In summary, we present three different methods to calculate the XAS integral and the OM. Each method has its advantages and disadvantages. The main message of these calculations is to show that the OM for vanadium in $VI_3$ is significantly large no matter which method we use.



In order to give a final number for the OM estimate which still describes the experimental problems in its determination, we decided to take the average value of the three methods proposed and use the standard deviation among them as the error bar, which comes down to and OM of **0.6(1) μ_B.**

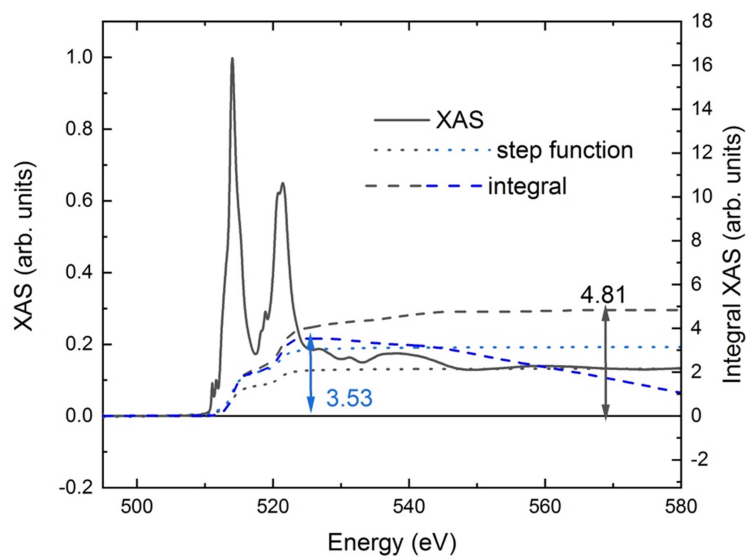

**Figure S5** - XAS measured at RT normal incidence (continuous line). Dotted line shows two different options for the step function: aligning with the XAS at 527 eV or at 570 eV. Dashed lines show the corresponding integral, following the color of the step function.

MULTIPLET SIMULATIONS



X-ray absorption spectra and x-ray magnetic circular dichroism spectra were simulated using ligand field multiplet simulations based on atomic Hartree-Fock calculations using Cowan's code[6] followed by chain symmetry proposed by Butler[7]. This approach has been used for the first time by Thole and van der Laan for the description of x-ray absorption spectra[2]. The local V symmetry used in the simulations was $D_{3d}$ which is reduced to $C_3$ when the magnetic moment is added. Slater Integrals were reduced by 50% from Hartree-Fock values. A crystal field splitting (CFS) $10D_q = 1.5$ eV was used. The trigonal distortion $D_\sigma$ was varied while $D_\tau$ was kept to zero. In the text, we refer to the size of trigonal distortion using the energy splitting of the $t_{2g}$ levels $E(e'_g) - E(a_{1g}) = 3D_\sigma + 20/3\,D_\tau = D_{trg}$. The $d$-SOC used was 13 meV following Lane et al.[8], which corresponds to 50% of the atomic value. The $p$-SOC was increased by 6% to fit with the experimental splitting of the $L_3$ and $L_2$ edges.

The energy levels in trigonal symmetry are related to the CFS parameters as shown in equations (2-4).

$$E(e') = -4 \times D_q + D_\sigma + {^2}\!/_3 \times D_\tau \qquad (2)$$

$$E(a_{1g}) = -4 \times D_q - 2 \times D_\sigma - 6 \times D_\tau \qquad (3)$$

$$E(e_g) = 6 \times D_q + {^7}\!/_3 \times D_\tau \qquad (4)$$



The choice of CFS parameters was based on DFT calculations and previous results, as from Lane et al.[8]

XAS, XMCD MEASUREMENTS

X-ray magnetic circular dichroism measurements were carried out using the EPFL-PSI end-station in the X-Treme beamline[9], Swiss Light Source. The crystals were mounted inside a glove box and cleaved for a fresh surface. Then the crystals were transferred in an inert atmosphere from the glove box to the load lock, where they were cleaved for a second time and then pumped down. To improve conductivity for the total electron yield measurements, the crystals were capped with a few nm carbon layer evaporated *in situ* in the ultra-high vacuum sample preparation chamber. The measurements shown were done with the x-rays perpendicular to the exfoliation plane (parallel to the *c*-axis in rhombohedral structure, $c^R$).

Figure S6 shows two measurements done during the same beamtime with a time difference of 6.5 hours. These measurements show that no sign of change in the XAS nor XMCD is observed due to long incidence time of the x-rays.



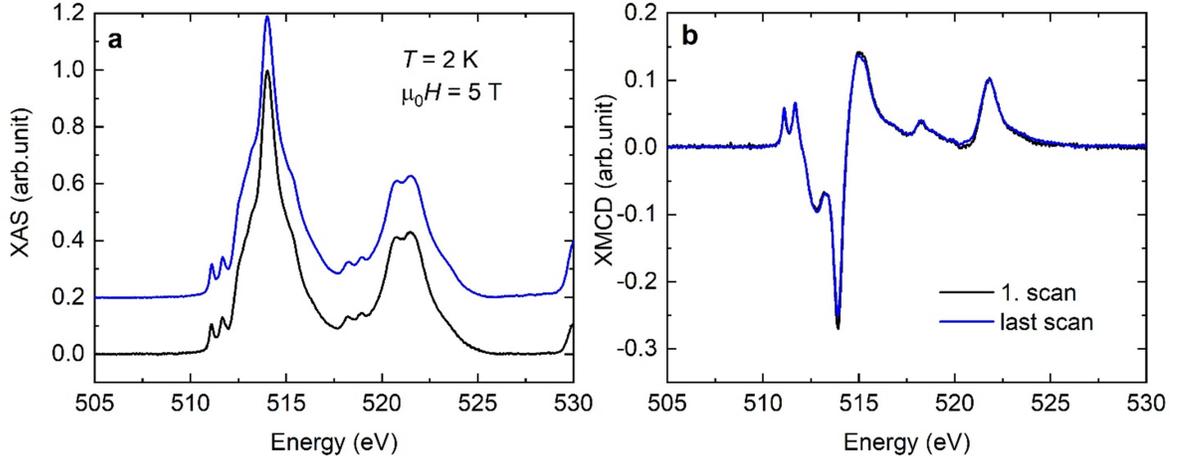

**Figure S6** - a) XAS and b) XMCD measurements done in the same crystal 6.5 hrs. apart in time showing no sign of radiation damage.

DFT CALCULATIONS

Density functional theory (DFT) calculations employed the full-potential linear augmented plane wave (FP-LAPW) method, as implemented in the band structure program ELK[10]. The generalized gradient approximation (GGA) parametrized by Perdew-Burke- Ernzerhof[11] has been used to determine the exchange-correlation potential. SOC is known to play a key role in the formation of OM and has been included in the calculation. DFT simulations utilizing GGA-PBE have already successfully described quasi-2D compound VI3[12]. The full Brillouin zone has been sampled by $10 \times 10 \times 5$ k-points and the convergence w.r.t. k-mesh density has been verified. An increased accuracy of expansion into spherical harmonics has been used with $l_{max}$=14. Since the material is known to be a Mott insulator, we have included the effect of



electron-electron correlations in terms of the Hubbard correction term U = 4.3 eV [13,14] together

with Hund's rule exchange parameter J = 0.8 eV, acting on $3d$ electrons of V. Double counting

was treated in the fully localized limit.